\documentclass[preprintnumbers, prd, onecolumn, floatfix, superscriptaddress, nofootinbib] {revtex4-2}
\usepackage{setspace}
\usepackage{graphicx}
\usepackage{subfig}
\usepackage{epsfig}
\usepackage{bm}
\usepackage{amssymb}
\usepackage{float}
\usepackage{amsmath}
\usepackage{dcolumn}
\usepackage[colorlinks]{hyperref}
\usepackage[usenames,dvipsnames]{color}
\usepackage{enumitem}
\hypersetup{ breaklinks=true, pdfstartview={FitH},
	colorlinks=true, linkcolor=blue, citecolor=red,
	filecolor=magenta, urlcolor=blue, anchorcolor=green,
	linktocpage=true }

\def\O{\Omega}

\def\Lam{\Lambda}
\def\doi{http://doi.org}

\def\d{$}
\def\r{\ref}
\def\and{$and$}
\def\g{~}
\def\a{\alpha}
\def\b{\beta}
\def\c{\gamma}
\def\l{\label}

\newcommand{\be}{\begin{equation}}
	\newcommand{\ee}{\end{equation}}
\newcommand{\ban}{\begin{eqnarray*}}
	\newcommand{\ean}{\end{eqnarray*}}
\newcommand{\ba}{\begin{eqnarray}}
	\newcommand{\ea}{\end{eqnarray}}
\newcommand{\no}{\nonumber}

\newcommand{\bc}{\begin{center}}
	\newcommand{\ec}{\end{center}}	
\newcommand{\bi}{\bibitem}
\usepackage{xcolor}

\begin{document}
	
\title{An Accelerating Flat FLRW Model with Observation Constraints and Dynamic $\Lambda$ }
	\author{G. K. Goswami}
	\email{gk.goswami9@gmail.com}
	\affiliation{Department of Mathematics, Netaji Subhas University of Technology, New Delhi-110 078, India}

	\author{Anirudh Pradhan}
	\email{pradhan.anirudh@gmail.com}
	\affiliation{Centre for Cosmology, Astrophysics and Space Science (CCASS), GLA University, Mathura-281 406, Uttar Pradesh, India}
	
	\begin{abstract}
		
		\begin{singlespace}
			\noindent In this paper, we explore power law solution of FLRW universe model that is associated with a variable cosmological term $\Lambda(t)$ as a linear function of $\frac{\ddot{a}}{a}, (\frac{\dot{a}}{a})^2$ and $\rho$. The model parameters were estimated on the basis of the four data sets: The Hubble 46 data, the Union 2.1 compilation data sets comprising of distance modulus of 580 SNIa supernovae at different redshifts, the Pantheon  data set which contains Apparent magnitudes of 1048 SNIa supernovae at various redshifts and finally BAO data set of  volume averaged distances at 5 redshifts. 
			We employ the conventional Bayesian methodology to analyze the observational data and also the Markov Chain Monte Carlo (MCMC) technique to derive the posterior distributions of the parameters. 
            The  best fit values of Hubble parameter $H_0$ as per the four data sets are found as $61.53^{+0.453}_{-0.456}$, $ 69.270^{+0.229}_{-0.228}$,
            $78.116^ {+0.480}_{-0.479}$, and $ 71.318 ^{+2.473}_{-2.283}$ respectively.
            Off late  the present value of Hubble parameters $H_0$ were empirically given as 73 and 67.7 (km/s)/Mpc using  distance ladder techniques and measurements of the cosmic microwave background. The OHD+BAO+Union and ~OHD+Pan+BAO+Union combined data sets provide the best fit Hubble parameter value $H_0$ as $67.427^{+0.197}_{-0.199}$ and $74.997^{+0.143}_{-0.145}$ respectively.
            The various geometrical and physical properties of the model were also investigated and were found in good agreements with observations. 
			
		\end{singlespace}
		
	\end{abstract}
	
\maketitle

	\section{Introduction}
	The expansion of the universe is speeding up rather than slowing
	down, according to cosmic observations from supernova type Ia
	(SNeIa) \cite{ref1,ref2,ref3}, the Cosmic
	Microwave Background (CMB), large scale
	structure,  weak lensing \cite{ref4,ref5,ref6,ref7}, Baryon 
        Acoustic Oscillation and  Spectroscopic
	Survey (BOSS) collaboration \cite{ref8, ref9}, Planck 
        collaboration \cite{ref10} and Actacama
	Cosmology Telescope Polarimeters (ACTPol) collaboration
	\cite{ref11}. It is a well-established fact that the majority
	of known gravitational phenomena can be explained by the
	successful general theory of relativity.
	Inclusion of cosmological constant in Einstein’s
		field equation gained importance as a positive cosmological
		constant is considered as a source of dark energy. $\Lambda$ CDM cosmology \cite{cop, gron} is just the Eddington–Lemaitre model with the difference that the cosmological constant term acts as a source of dark energy with equation of state $p_{\Lambda}= -\rho_{\Lambda.}$ Latest works by Abazajian et al. and Sahni and Starobinsky \cite{AbaAde, ref12} support $\Lambda$ CDM cosmology.
	A dynamic cosmological term $\Lambda(t)[$\cite{vish,azri,Hazri,szy,kous}] continues to be of interest in contemporary cosmology theories because it provides a natural solution to the cosmological constant problem. Significant observational evidence points to the finding of either $\Lambda$ or a component of the universe's material content that behaves
	similarly to $\Lambda$ and fluctuates slowly with time and space. 
	There is currently strong evidence from a variety of
	observations that the cosmos has a non-zero cosmological term
	\cite{ref13}. A cosmological term in the context of quantum 
        field theory is equivalent to the vacuum's energy density. It has been suggested that an excited vacuum fluctuation that sets off an inflationary expansion and super-cooling is what gave rise to the universe. The reheating that follows is caused by the release of trapped vacuum energy. A $\Lambda$, in the context of quantum field theory, is equivalent to the vacuum's energy density. It has been suggested that an excited vacuum fluctuation that sets off an inflationary expansion and super
	cooling is what gave rise to the universe. The reheating that
	follows is caused by the release of trapped vacuum energy. The
	$\Lambda$ acts as a repulsive force against the gravitational
	attraction between galaxies and is a measure of the energy of
	empty space. Since mass and energy are equal, if the $\Lambda$
	exists, the energy it represents counts as mass. Inflation could
	result from the energy of the $\Lambda$ plus the stuff in the
	universe if it is large enough. The universe with a cosmological
	term would expand more quickly over time because to the force
	emanating from the cosmological term, in contrast to normal
	inflation \cite{ref14}.
	Recent discussions by Dolgov \cite{ref15,ref16,ref17}, Sahni and
	Starobinsky \cite{ref12}, Ratra and Peebles \cite{ref18}, and
	others on the ``problem" of the cosmological constant and
	cosmology with a time-varying cosmological constant have noted
	that the cosmological constant is a ``constant" in the absence
	of any interaction with matter or radiation. Off late the authors\cite{ref19,ref20,ref21,ref22} have developed accelerating universe model in f(R,T) gravity theory which shows transition from  deceleration to acceleration and fits well on observational grounds.\\
	
	In this paper, we attempted to model a physical spatially homogeneous and isotropic flat  FLRW(Friedmann Lemaitre Robertson Walker) universe in  GR(General relativity) by considering variable time dependent cosmological constant(CC) $\Lambda$. On the basic of Einstein Field equations for FLRW space time, the CC is assumed as linear function of $\frac{\ddot{a}}{a}$ describing acceleration, $\frac{\dot{a}^2}{a^2}$ which is square of Hubble parameter describing rate of expansion and the density $\rho$ of the universe. On solving the field equations, we found that our model represents an accelerating Einstein-de Sitter universe. The model had undergone under the constraints of the four data sets. The Hubble 46 data set(OHD) describing Hubble parameter values at various redshifts(Table1 for references). Union 2.1 compilation data sets (Union) comprising of distance modulus of 580 SNIa supernovae at different redshifts \cite{ref23}. The Pantheon  data set(Pan) which contains Apparent magnitudes of 1048 SNIa supernovae at various redshifts \cite{ref24,ref25} and finally BAO data set(BAO) of  volume averaged distances at 5 redshifts \cite{ref26}. These data sets and their combinations were used to estimate the model parameters which also include $H_0$ the present value of Hubble parameters. The  best fit values of Hubble parameter $H_0$ as per the four data sets described earlier are found as $61.53^{+0.453}_{-0.456}$, $ 69.270^{+0.229}_{-0.228}$,
    $78.116^ {+0.480}_{-0.479}$ and $ 71.318 ^{+2.473}_{-2.283}$ respectively.
   While ``early universe" techniques based on cosmic microwave background measurements agree on a value of 67.7 (km/s)/Mpc, the calibrated distance ladder techniques have converged on a value of $H_0$ approximately 73 (km/s)/Mpc.
     We also tried to achieve these values statistically by using combined data sets out of the four described earlier. We also worked out in finding the  growth of matter and vacuum cosmological constant related energy densities of the universe. They do follow the increasing trend  over redshifts which interprets that in the past the densities were high and over the time they are gradually  decreasing due to the expansion of the universe. We have estimated that the present values  $\rho_0 $ and $\rho_{\Lambda_0} $ of matter and Lambda vacuum densities are $3.244*10^{-30}gm/cm^3$ and $7.39*10^{-30}gm/cm^3$ respectively. The higher value of vacuum energy shows the dominance of this energy at present. We recall that this energy is responsible for the present acceleration of the universe as it creates negative anti gravitating pressure in the universe. Finally We have also developed  the association of time over redshifts and estimated  the present age of the universe. As per our model it is estimated as 21.7282 billion yrs which is considerably higher than that predicted by Lambda CDM model.\\
    
    The paper is lined up in the following order. In the section 2, we present Einstein field equations for FLRW space time with variable cosmological constant. In this section we have also solved the field equations and obtained power law solutions for Hubble parameter, cosmological constant, matter and Lambda dominated  energy densities. Section 3 has five subsections in which we have estimated model parameters $H_0$, $\alpha$, $\beta$ and $\gamma$ with the help of four data sets with short names OHD, Union, Pan and BAO. We have also estimated by combining these data sets to get more better results which fit best with the observations. In section 4, we have obtained present values of matter, Lambda dominated vacuum energy density in proper units. In section 5, we have presented a expression to convert redshifts into time. we have also obtained present age of the universe. All sections are very well attached with figures and tables to look our findings at first sight. the last section contains some useful concluding remarks. 
     
\section{{Einstein Field equations with variable cosmological
	constant\g$\Lambda(t)$\g and Their Solutions}}\l{2}
	We begin with the Einstein's field equations with cosmological
	constant
	\be R_{\mu\nu}-\frac{R}{2}g_{\mu\nu}+\Lambda g_{\mu\nu}=8\pi G
	T_{\mu\nu},\label{efe}\ee
	where \d R_{\mu\nu},\g R,\g g_{\mu\nu} \g \and \g T_{\mu\nu}
	\d\g are the Ricci tensor, Ricci scalar, metric tensor and
	energy-momentum tensor of the contents of the universe.
	The Friedmann-Robertson-Walker (FRW) metric is given by the line
	element
	\be
	ds^2=dt^2-a^2(t)\left[\frac{dr^2}{1-kr^2}+r^2(d\theta^2+\sin^2\theta
	d\phi^2)\right],
	\label{fr}\ee 
	where the curvature parameter $k=-1,0,1$ for open, flat and
	closed models of the universe, respectively and $a(t)$ is the
	scale factor which geometrically expands the universe over time.
	Since current observations from CMB detectors such as BOOMERanG,
	MAXIMA, DASI, CBI and WMAP confirm a spatially
	flat universe, $k=0$.
	We consider universe to be filled with perfect fluid whose
	energy momentum tensor \d T^\mu_\nu\d is given as
	\be
	T^\mu_\nu = (p + \rho)u^\mu u_\nu - p \delta^\mu_\nu,
	\label{emt}\ee
	where $ p $ and $ \rho $ are the pressure and matter-energy
	density of the perfect fluid. We have considered velocity vector
	$ u^i = (1,0,0,0) $, so that $ u^i u_i = 1 $. Velocity of light
	$c$ is taken as 1.
	Solving Einstein's Field equations for FRW metric, we get
	following equations as
	\be
	2\frac{\ddot{a}}{a}+\frac{\dot{a}^2}{a^2}-\Lambda =-8\pi G p,
	\label{freq1}\ee
	\be
	\frac{\dot{a}^2}{a^2}-\frac{\Lambda }{3}=\frac{8\pi G}{3}\rho.
	\label{freq2}\ee
    Here, dot denotes the time derivative. We want to develop an universe model with variable cosmological
	constant with a view to geometries gravity. It may result in
	getting an accelerating universe. The continuity equation of total fluid (matter with energy density $\rho$ and vacuum energy density $\Lambda$) will be written as $\dot{\rho}+3H(\rho+p)+\dot{\Lambda}=0$, where $\rho$ and $\Lambda$ are varying with time. Here, we take for the time being $p=\gamma\rho$ and $\Lambda$ denotes the varying vacuum energy satisfying $\omega_\Lambda=-1$. For $\gamma=0$, one may have the pressure-less dust and for $\gamma=\frac{1}{3}$, the fluid will be radiation. For $\gamma=1$, the stiff fluid may be visualized in $p=\gamma\rho$ relation. From the field equations, we
	infer that the dynamic cosmological term $\Lambda(t)$ is
	possibly a linear function of $\frac{\ddot{a}}{a},
	(\frac{\dot{a}}{a})^2$ and $\rho$ as follows:
	\be
	\Lambda(t) = \a\frac{\ddot{a}}{a}+\b
	\left(\frac{\dot{a}}{a}\right)^2+4\pi G\c\rho,
	\label{lamb}\ee 
	where \d \a,\g \b \g \d are arbitrary constants and $\gamma$ is equation of state parameter of fluid. In the present analysis, we consider $\gamma$ as a parameter and estimate its value from the observational data. On solving Eqs.(\r{freq1}), (\r{freq2}) and (\r{lamb}), we get
	the following expression for Hubble parameter \d H \d,\g
	\d\rho\d \g and \g \d \Lambda \d.
	
	\be
	H(t) = \frac{H_0 (-\c +\a (w+1)-2)}{H_0 (t-t_0) (w+1) (\b
		+\a-3)+(-\c +\a (w+1)-2)},
	\l{Hub}\ee
	
	\be
	\rho (t)=\frac{{H_0}^2 (\a +\b -3) (\a -\c +\a w-2)}{4 \pi G (\a
		-\c +H_0(t-t_0) (w+1) (\a +\b -3)+\a w-2)^2},
	\l{rho}\ee
	
\be
p (t)=\frac{\omega{H_0}^2 (\a +\b -3) (\a -\c +\a w-2)}{4 \pi G (\a
	-\c +H_0(t-t_0) (w+1) (\a +\b -3)+\a w-2)^2},
\l{pres}\ee

	\g\g\g where we have considered equation of state for perfect
	fluid as \d p = w \rho \d\g and boundary condition is taken as
	\d H=H_0 \d \g when\g \d t= t_0. \d
	
	We define energy parameters \d \O _m(t)\g \and\g \O _{\Lam }(t)
	\d\g as follows:
	
	\be
	\O _m(t){=}\frac{8 \pi G \rho (t)}{3 H(t)^2} = \frac{2 (\alpha
		+\beta -3)}{3 (\alpha -\gamma +\alpha w-2)}
	\l{Om_m}\ee
	and
	\be
	\O _{\Lam }(t){=}\frac{\Lam (t)}{3 H(t)^2} =\frac{\alpha -2
		\beta -3 \gamma +3 \alpha w}{3 (\alpha -\gamma +\alpha w-2)}
	\l{Om_l}\ee
	So that from Eq. (\r{freq2}), we get
	\be\no
	\O _m(t) + \O _{\Lam }(t) =1
	\ee
	It is verified that in our model too, the above energy relation
	holds good.\\
	Integrating Eq.(\r{Hub}), we get expression for scale factor as
	follows:
	\be
	a(t)= a_0 \left(\frac{-\c +H_0 (w+1) (\a +\b -3) (t-t_0)+\a
		(w+1)-2}{-\c +\a (w+1)-2}\right)^{\frac{\a -\c +\a w-2}{(w+1)
			(\a +\b -3)}},
	\l{sca}\ee
	where  boundary condition is taken as 
	\d a=a_0 \d \g when\g \d t= t_0. \d\\
	From Eqs.(\r{Hub}) and (\r{sca}), we get a very familiar
	relation between Hubble para meter\g \d `H'\d \g and scale
	factor\g \d `a'\d \g as follows:
	\be \no
	H(t)={H_0} \left(\frac{a_0}{a(t)}\right)^\frac{(w+1) (\alpha
		+\beta -3)}{-\gamma +\alpha (w+1)-2};
	\ee
	Now using one more familiar relation between scale factor $a$
	and redshift $z$ as:
	\be\no 
	\left(\frac{a_0}{a(t)}\right)= 1+z,
	\ee
	we get following a very important relation:
	\be 
	H(z)=H_0 (1+z)^\frac{(w+1) (\alpha +\beta -3)}{-\gamma +\alpha
		(w+1)-2};
	\l{Hubz}\ee
	The deceleration and jerk parameters \d q \d \g and \g \d j \d\g
	are obtained from the following formalism:
		\begin{equation}
		q:= -\frac{a\ddot{a}}{\dot{a}^{2}}=\frac{d}{dt}\left( 
		\frac{1}{H}\right) -1=\frac{d\ln H}{d\ln (1+z)}-1,
	\end{equation}%
	and 
	\begin{equation}
		j:= \frac{a^{2}\dddot{a}}{\dot{a}^{3}}=
		q\left(2q+1\right)-\frac{\dot{q}}{H}=q\left(2q+1\right)+\frac{dq}{d\ln
			(1+z)}.\label{jerk}
	\end{equation}
	In our model, these parameters are obtained as 
	\be
	q(z)=\frac{(w+1) (\alpha +\beta -3)}{-\gamma +\alpha (w+1)-2}-1,
	\l{dec}\ee 
	and
	\be 
	j(z)=\frac{(\beta +\gamma +(\beta -3) w-1) (\alpha +2 \beta
		+\gamma +w (\alpha +2 \beta -6)-4)}{(\alpha -\gamma +\alpha
		w-2)^2}
	\l{jerk}\ee
	The density \g\d\rho\d\g of the universe and the cosmological
	constant \g\d\Lam\d\g are obtained from Eqs.(\r{Om_m}) and
	(\r{Om_l}) as:
	\be 
	\rho(z) = \rho_c \frac{2(\alpha +\beta -3) (z+1)^{\frac{2 (w+1)
				(\alpha +\beta -3)}{-\gamma +\alpha (w+1)-2}}}{3 (\alpha -\gamma
		+\alpha w-2)}
	\l{dens}\ee
	and
	\be
	\Lambda(z)=\frac{{H_0}^2 (\alpha -2 \beta -3 \gamma +3 \alpha w)
		(z+1)^{\frac{2 (w+1) (\alpha +\beta -3)}{-\gamma +\alpha
				(w+1)-2}}}{\alpha -\gamma +\alpha w-2},
	\l{Lamb}\ee
	where $\rho_c= \frac{3H_0^2}{8\pi G}$ is the critical density of
	the universe.\\
	From these equations, we may obtained the present values of
	density and cosmological constant as
	\be\no 
	\rho_0= \rho_c \frac{2(\alpha +\beta -3)} {3(\alpha -\gamma
		+\alpha w-2)}
	\ee
	and
	\be\no 
	\Lambda_0=\frac{{H_0}^2 (\alpha -2 \beta -3 \gamma +3 \alpha
		w) }{\alpha -\gamma +\alpha w-2}.
	\ee
	So, we may express \g\d\rho(z)\d\g and \g\d\Lambda(z)\d\g in
	term of their present values as:
	\be	
	\rho(z)=\rho_0 (z+1)^{\frac{2 (w+1) (\alpha +\beta -3)}{-\gamma
			+\alpha (w+1)-2}}
	\l{rho}\ee
	and
	\be
	\Lambda(z)=\Lambda_0 (z+1)^{\frac{2 (w+1) (\alpha +\beta
			-3)}{-\gamma +\alpha (w+1)-2}}.
	\l{Lam}\ee
	These equations indicate that our model resembles
	Einstein-de Sitter universe with the difference that it describes
	accelerating universe with variable power law oriented
	cosmological constant provided that the deceleration constant given by Eq. (\r{dec}) is positive. It will be seen from the latter sections that it is indeed positive. We may comment that our model present the current scenario of the accelerating universe. 
	\section{Estimations of model parameters $\alpha$, $\beta$, $\gamma$, and 
		$H_0$ from various available SNIa datasets, Hubble datasets and BAO datasets.}
	\subsection{Estimations of Model parameters from Hubble
		dataset.}\l{A}
    We consider the  Hubble parameter table~\cite{Bhardwaj} consisting of
    46 data set of of observed values of ~ $H$   ~ for various
    redshift in the range ~ $ 0 \leq z \leq 2.36 $ ~ with
    possible error in observations.\\
	We use this data set to estimate the model parameters \g\d
	\a,\b,\c,\g \and\g H_0 \d\g in multiple ways. First we fit
	Hubble parameter function given by Eq. (\r{Hubz}) to these data
	by method of least square and estimates model parameters. There
	after assuming these estimated values as initial guess, we use
	method of least $\chi^{2}$ to make more refined estimations.
	Finally we carry Markov Chain Monte Carlo (MCMC) method to further
	refine the estimations by assuming  $\chi^{2}$ estimations as
	initial guess. We recall that $\chi^{2}$ formula for Hubble
	function is as follows:
	\be
	\chi^{2}(\a,\b,\c,H_0) =
	\sum\limits_{i=1}^{46}\frac{(H_{th}(z_{i},\a,\b,\c,H_0) -
		H_{ob}(z_{i}))^{2}}{\sigma {(z_{i})}^{2}}.
	\l{chi}\ee
	Our estimations are shown in the following Table-2
	\begin{table}[H]\l{table2}
		\centering
		{\begin{tabular}{c|c|c|c|}
		\hline\hline
				Parameters & Least square & Least\g\d \chi^2\d\g & MCMC
				simulation\\
				& Estimations & Estimations & Estimations\\
				\\
				\hline
				\\
				\d\a\d & $0.524$ & $1.235\pm 0.024$ & $1.234\pm0.024$\\
				\\
				\hline
				\\
				\d\b\d & $1.91$ & $0.816\pm0.024$ & $0.817\pm0.024$\\
				\\
				\hline
				\\
				\d\c\d & $0.908$ & $0.685\pm 0.008$ & $0.685\pm0.008$\\
				\\
				\hline
				\\
				
				\d H_{0}\d & $59.609$ & $61.537 \pm 0.461$ &
				$61.53^{+0.453}_{-0.456}$\\
				\\
				\hline\hline
		\end{tabular}}
		\caption{\small{ The best fit values of the model parameters
				$\a$, $\b$, $\c$ and $H_{0}$ for the best fit $H(z)$ Hubble
				curve.}}
	\end{table}\l{table2}
	Now we substantiate our work by presenting various figures in
	form of plots. Fig.1 describes the Hubble parameter $H(z)$
	versus redshifts $z$ plot and error bar plot for the best fit
	values of model parameters g$\alpha$, $\beta$, $\gamma$ and $H_{0}$
	using methods of least square and minimum $ \chi^2$ 
	function value. Fig.2 describes MCMC Simulation based
	estimations for model parameters g$\alpha$, $\beta$, $\gamma$ and $H_{0}$.
	\begin{figure}[H]
		\centering
		
		a.\includegraphics[width=6cm,height=6cm,angle=0]{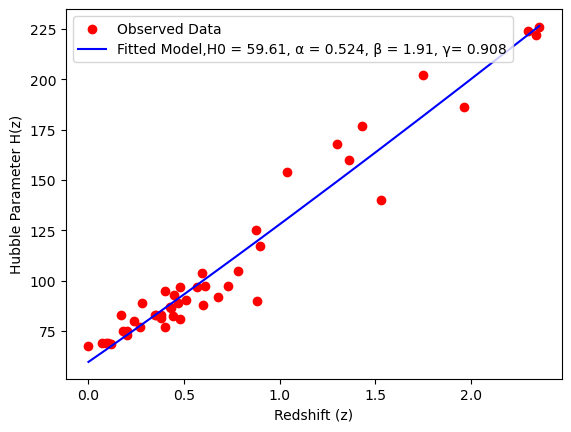}
		b.\includegraphics[width=6cm,height=6cm,angle=0]{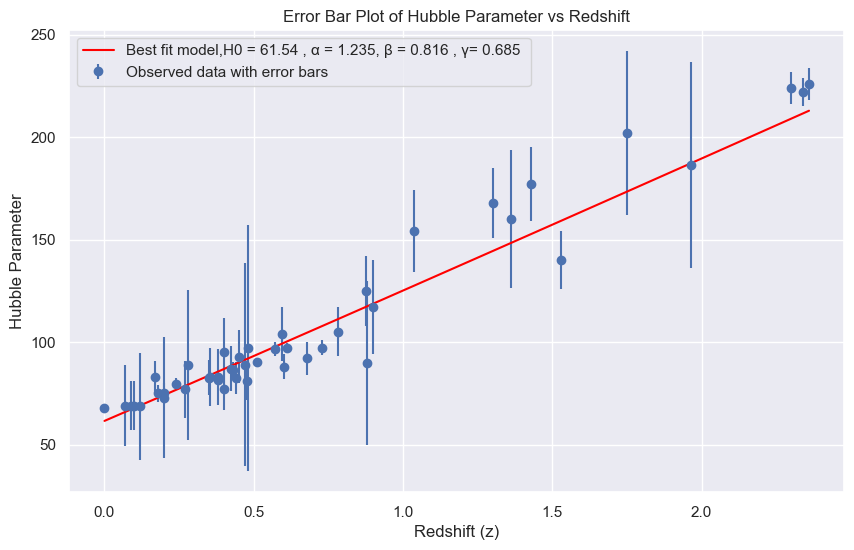}
		\caption{The Hubble parameter $ H(z) $ versus redshifts $z$
			Plot and Error bar Plot for the best fit values of model
			parameters  $\alpha$, $\beta$, $\gamma$ and $H_{0}$ using method of
			least square and minimum $\chi^2 $ function value.}
	\end{figure}\l{fig1}
	
	\begin{figure}[H]
		\centering
		\includegraphics[width=6cm,height=6cm,angle=0]{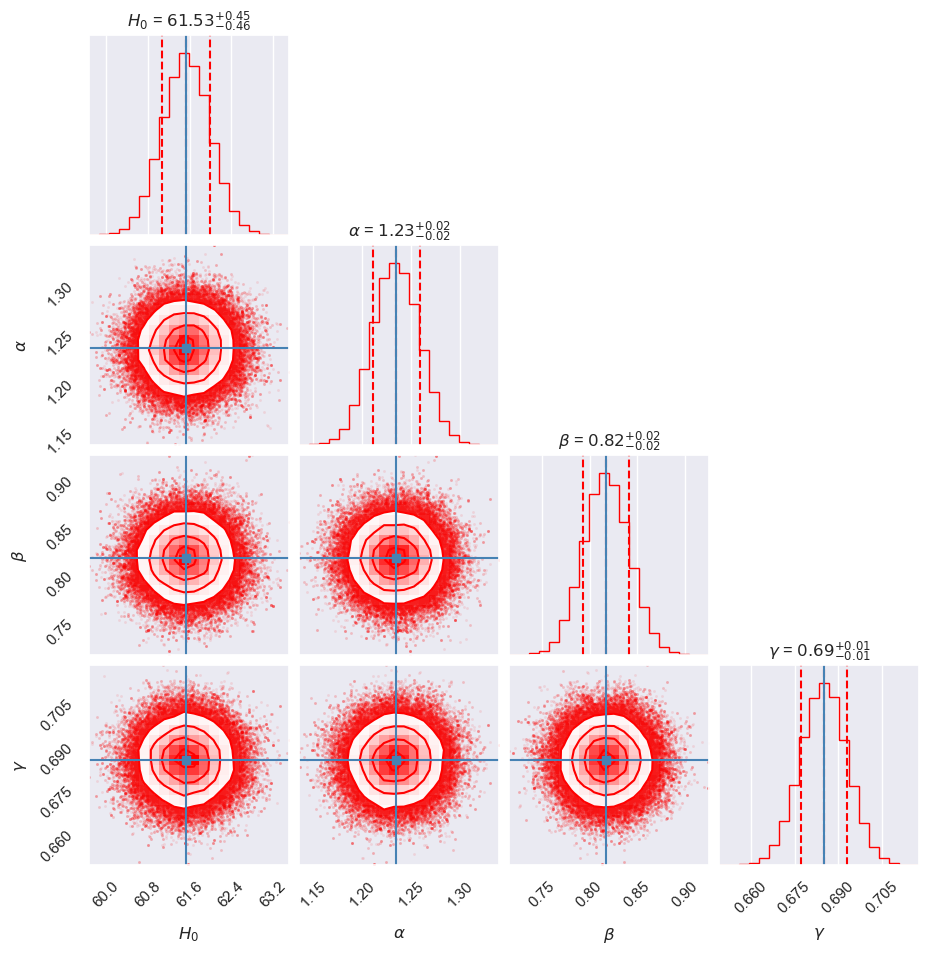}
		\caption{Markov Chain Monte Carlo (MCMC) Simulation based estimations for Model
			parameters $\a$, $\b$, $\c$ and $H_{0}$ and Corner plots for 36
			Hubble data set.}
	\end{figure}\l{fig2}

	\subsection{Estimations of Model parameters from Supernova SNIa Union 2.1 Compilation 680 data set.}\l{B}
	The Supernova SNIa Union 2.1 Compilation data set is comprised
	of 650 data set of Distance modulus of SNIa supernovae for
	various redshifts in the range \d 0\leq z \leq 1.4 \d \ associated
	with observational errors. The theoretical formula for distance
	modulus is given as:
	\begin{equation}
		\mu_{th}(z;h,\boldsymbol{params}):= 5 \log\left[
		D_L(z;\boldsymbol{params})\right] +\mu_0(h) . \label{mu}
	\end{equation}
	Here $h$ is the Hubble constant in units of $100$ km/s
	Mpc$^{-1}$, $\boldsymbol{params}$ denotes the set of
	cosmological parameters of interest other than $h$. In our model
	params are \g\d\a,\b\g \and\g \c \d.
	\begin{equation}\no
		\mu_0(h):= 5\log\left(\frac{10^3 c/(\mbox{km/s})}{h}\right) =
		42.38-5\log h.
	\end{equation}
	and
	\begin{equation}\no
		D_L(z;h,\boldsymbol{params})=(1+z)\int_0^z
		{\frac{dz'}{H(z';h,\boldsymbol{params})/H_0}}\;,
	\end{equation}
	is the luminosity distance. We use Hubble function as per
	Eq.(\r{Hubz}). \\
	Like in sub section (a),we use union 2.1 data set to estimate
	the model parameters \g\d \a,\b,\c,\g \and\g H_0 \d\g in
	multiple ways. First we fit distance modulus \g(\d m_{u}\d)\g
	function given by Eq.(\r{mu}) to observational distance modulus
	\g(\d m_{u}\d)\g data by method of least square and estimates model
	parameters. There after assuming these estimated values as
	initial guess, we use method of least $\chi^{2}$ to make more
	refined estimations.
	Finally we carry Markov Chain Monte Carlo (MCMC) method to further
	refine the estimations by assuming $\chi^{2}$ estimations as
	initial guess. We recall that $\chi^{2}$ formula for distance
	modulus \g(\d\mu\d)\g is as follows:
	\be
	\chi^{2}(\a,\b,\c,H_0) =
	\sum\limits_{i=1}^{580}\frac{(m_{bth}(z_{i},\a,\b,\c,H_0) -
		m_{bob}(z_{i}))^{2}}{\sigma {(z_{i})}^{2}}.
	\l{chi}\ee	
	Our estimations are shown in the following Table-3
	\begin{table}[H] \l{table3}
		\bc
		{\begin{tabular}{c|c|c|c|}
				\hline\hline 
				Parameters & Least square & Least\g\d \chi^2\d\g & MCMC
				simulation\\
				& Estimations & Estimations & Estimations\\  
				\hline\\
				\d\a\d & $1.112$ & $1.115\pm 0.03$ & $ 0.987  \pm 0.030$\\\\				\hline\\
				\d\b\d & $ 0.944$ & $ 0.94\pm 0.03$ & $ 0.961 \pm 0.030$\\\\				\hline\\
				\d\c\d & $ 0.943$ & $ 0.941\pm 0.01$ & $0.866\pm0.01$\\\\
				\hline\\
				\d H_{0}\d & $68.694$ & $69.273 \pm 0.23$ & $ 69.270^{+0.229}_{
					-0.228}$\\\\
				
				\hline\hline
		\end{tabular}}
		\ec
		\caption{\small{ The best fit values of the model parameters
				$\a$, $\b$, $\c$ and $H_{0}$ for the best fit distance modulus
				\g\d \mu(z)\d\g curve.}}
	\end{table}\l{table3}
	
	Now we substantiate our work by presenting various figures in
	form of plots. Fig.4 describes the distance modulus \d \mu(z) \d
	versus redshifts $z$ Plot and Error bar Plot for the best fit
	values of model parameters \g$\a$, $\b$, $\c$ and $H_{0}$\g
	using methods of least square and minimum\g \d \chi^2 \d \g
	function value. Fig.5 describes Markov Chain Monte Carlo (MCMC) simulation based
	estimations for Model parameters $\a$, $\b$, $\c$ and $H_{0}$.
	
	\begin{figure}[H]
		\bc
		
		a.\includegraphics[width=6cm,height=6cm,angle=0]{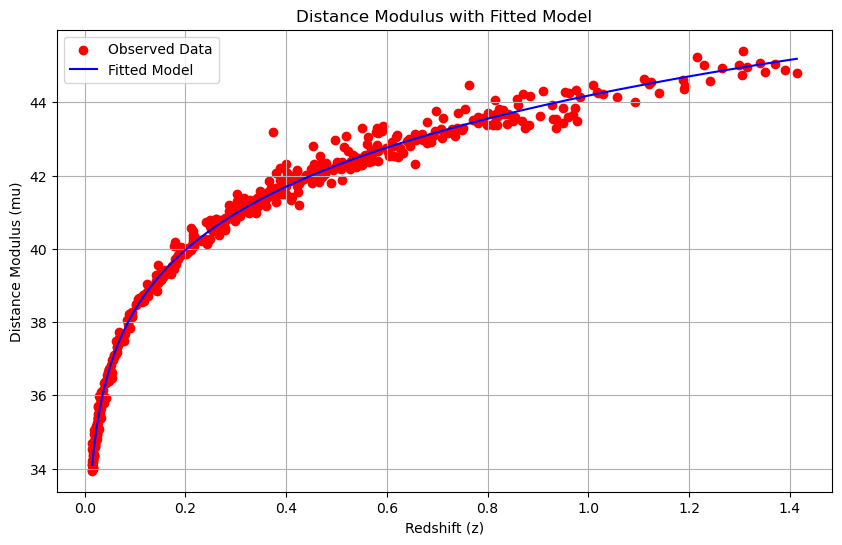}
		b.\includegraphics[width=6cm,height=6cm,angle=0]{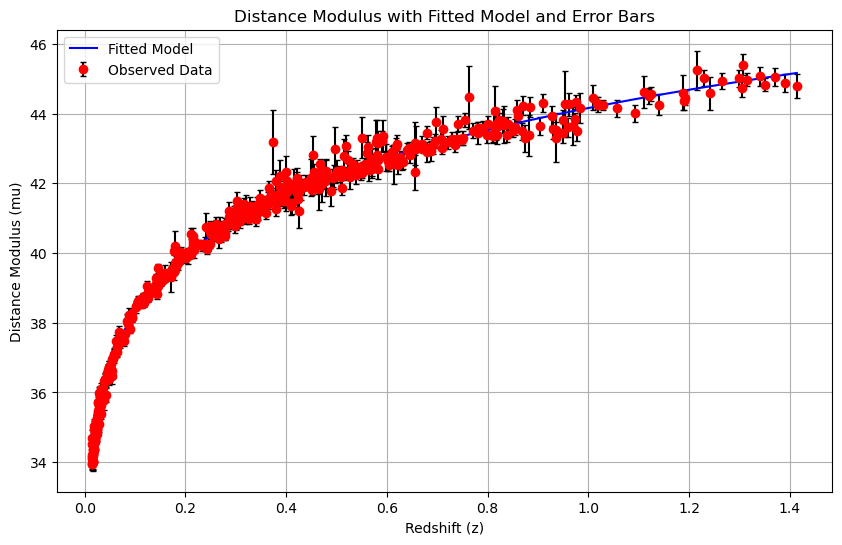}
		\ec
		\caption{The distance modulus \g(\d\mu\d)\g versus redshifts $z$
			Plot and Error bar Plot for the best fit values of model
			parameters \g$\a$, $\b$, $\c$ and $H_{0}$\g using method of
			least square and minimum\g \d \chi^2 \d \g function value .
			Estimated values are, \d \g \a =0.987\pm 0.03, \b = 0.961\pm
			0.03,\c =0.866\pm0.01\d and $69.270 \pm 0.23\d.}
	\end{figure}\l{fig4}
	
	\begin{figure}[H]
		\bc
		\includegraphics[width=6cm,height=6cm,angle=0]{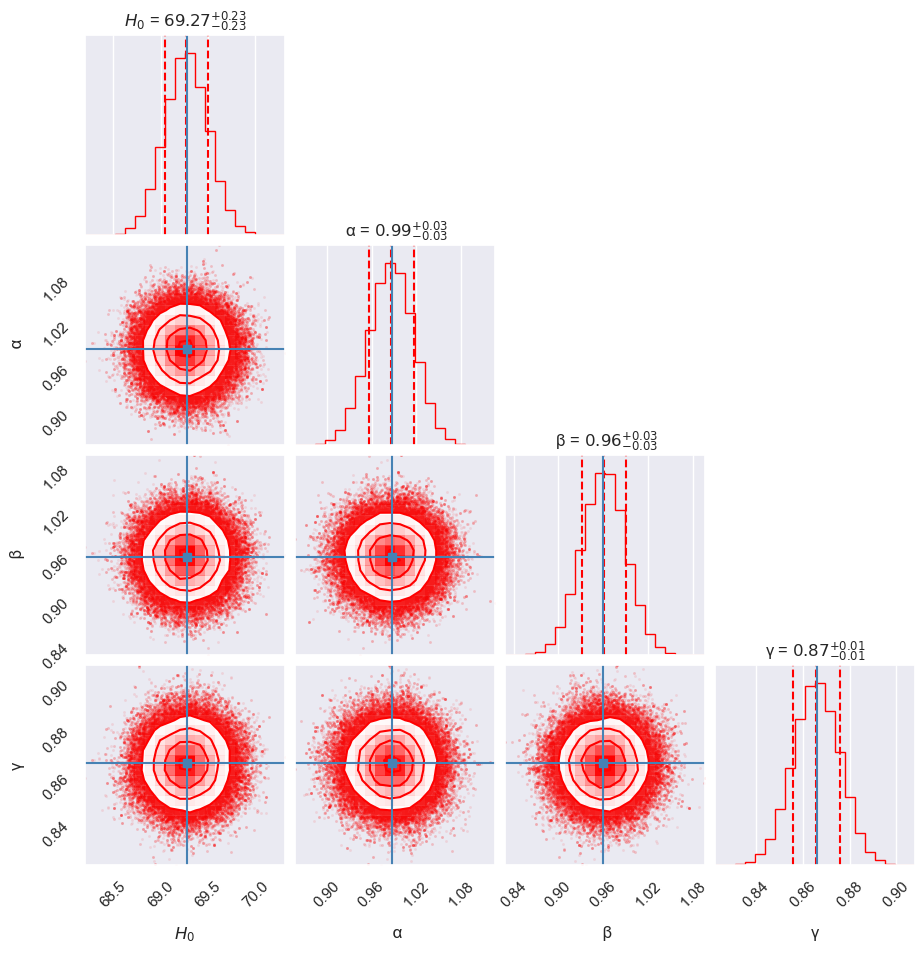}
		\ec
		\caption{Markov Chain Monte Carlo (MCMC) Simulation estimations for Model
			parameters $\a$, $\b$, $\c$ and $H_{0}$ and Corner plots for
			Union 2.1 SNIa 680 data set . }
	\end{figure}\l{fig5}

	\subsection{Estimations of Model parameters from Supernova SNIa
		1048 Pantheon data set.}\l{C}
	The Supernova SNIa Pantheon data set is comprised of 1048 data
	set of Apparent magnitude \g(\d m_b\d)\g of SNIa supernovae for
	various redshifts in the range \g\d 0\leq z \leq 2.24\d \g,
	associated with observational errors. The theoretical formula
	for Apparent magnitude \g(\d m_b\d)\g is given as:
	
	\begin{equation}\no
		m_b = M + \mu(z)
	\end{equation}
	where $ M $ is the absolute magnitude of the object and $ \mu $
	is the distance modulus. We recall that Type Ia supernovae are
	considered ``standard candles" because they have a relatively
	uniform intrinsic brightness, which allows astronomers to use
	them to measure distances accurately. The absolute magnitude of
	a typical Type Ia supernova at its peak brightness is
	approximately:
	Absolute Magnitude: M = -19.09 in the B- band(blue light)\cite{cop}
	\begin{equation}
		m_{bth}(z;h,\boldsymbol{params}):= 5 \log\left[
		D_L(z;\boldsymbol{params})\right] + m_{b0}(h) .
		\label{mb}\end{equation}
	Here $h$ is the Hubble constant in units of $100$ km/s
	Mpc$^{-1}$, $\boldsymbol{params}$ denotes the set of
	cosmological parameters of interest other than $h$, in our model
	params are \g\d\a,\b\g \and\g \c \d.
	\begin{equation}\no
		m_{b0}(h):= 5\log\left(\frac{10^3 c/(\mbox{km/s})}{h}\right)
		-19.09 = 23.29-5\log h.
	\end{equation}
	and
	\begin{equation}\no
		D_{L}(z;h,\boldsymbol{params})=(1+z)\int_0^z
		{\frac{dz'}{H(z';h,\boldsymbol{params})/H_0}}\;,
	\end{equation}
	is the luminosity distance. We use Hubble function as per
	Eq.(\r{Hubz}). \\
	Like in sub section (a) and (b), we use Pantheon data set to
	estimate the model parameters \g\d \a,\b,\c,\g \and\g H_0 \d\g
	in multiple ways. First we fit Apparent magnitude \g(\d m_b\d)\g
	function given by Eq.(\r{mu}) to observational Apparent
	magnitude \g(\d m_{b}\d)\g data by method of least square and
	estimates model parameters. There after assuming these estimated
	values as initial guess, we use method of least $\chi^{2}$ to
	make more refined estimations.
	Finally we carry Markov Chain Monte Carlo (MCMC) simulations method to further
	refine the estimations by assuming $\chi^{2}$ estimations as
	initial guess. We recall that $\chi^{2}$ formula for Apparent
	magnitude \g(\d m_{b}\d)\g is as follows:
	\be
	\chi^{2}(\a,\b,\c,H_0) =
	\sum\limits_{i=1}^{1048}\frac{(m_{b}th (z_{i},\a,\b,\c,H_0)
 - m_{b}ob (z_{i}))^{2}}{\sigma {(z_{i})}^{2}}.
	\l{chi}\ee	
	Our estimations are shown in the following Table-4
	\begin{table}[H]\l{table4}
		\centering
		{\begin{tabular}{c|c|c|c|} 
				\hline\hline 
				Parameters & Least square & Least\g\d \chi^2\d\g & MCMC
				simulation\\
				& Estimations & Estimations & Estimations\\ 
				\hline\\ 
				\d\a\d & $1.145$ & $ 1.522\pm 0.49$ & $ 1.521^ {+0.488} _
				{-0.485}$\\\\
				\hline\\
				\d\b\d & $0.928$ & $ 0.880 \pm 0.49$ & $ 0.083^ {+0.490} _
				{-0.479}$\\\\
				\hline\\
				\d\c\d & $ 0.927$ & $ 1.163 \pm 0.49$ & $ 1.156^ {+0.490} _
				{-0.479}$\\\\
				\hline\\
				\d H_{0}\d & $ 77.713$ & $78.122 \pm 0.49$ & $78.116^ {+0.480} _
				{-0.479}$\\\\
				
				\hline\hline
		\end{tabular}}
		\caption{\small{ The best fit values of the model parameters
				$\a$, $\b$, $\c$ and $H_{0}$ for the best fit Apparent magnitude
				\g\d m_{b}(z)\d\g curve.}}
	\end{table}\l{table4}
	Now we substantiate our work by presenting various figures in
	form of plots. Fig.7 describes the distance modulus \d \mu(z) \d
	versus redshifts $z$ Plot and Error bar Plot for the best fit
	values of model parameters \g$\a$, $\b$, $\c$ and $H_{0}$\g
	using methods of least square and minimum $\chi^2$ 
	function value. Fig.8 describes Markov Chain Monte Carlo (MCMC) Simulation based
	MCMC estimations for Model parameters $\a$, $\b$, $\c$ and
	$H_{0}$.
	\begin{figure}[H]
		\centering
		
		a.\includegraphics[width=6cm,height=6cm,angle=0]{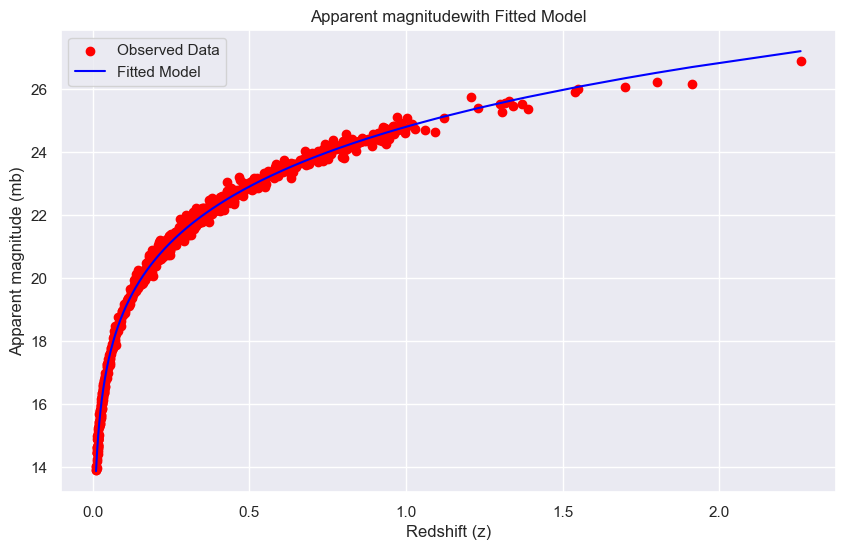}
		b.\includegraphics[width=6cm,height=6cm,angle=0]{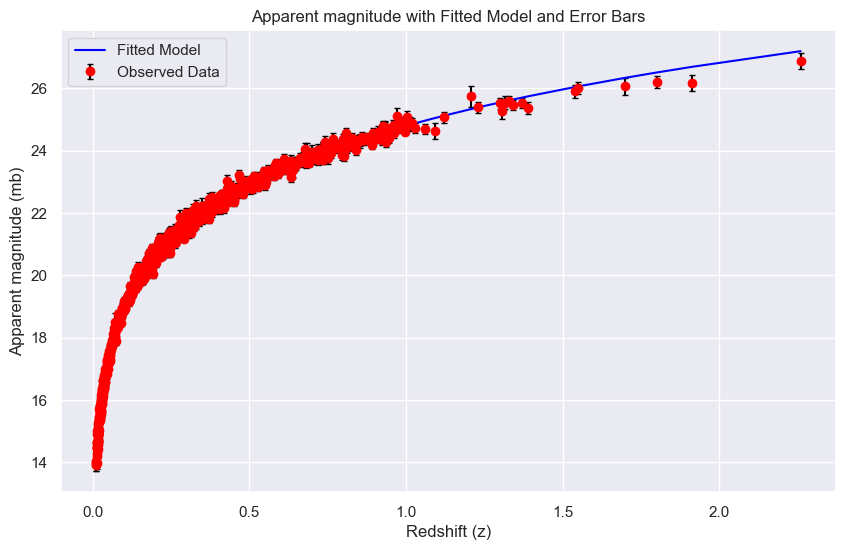}
		\caption{The Apparent magnitude of least square and minimum 
			$\chi^2$  function value. Estimated values are \d H_0 =
			78.122 \pm 0.49, \a =1.522\pm 0.49, \b = 0.880\pm 0.49,\c =
			1.163 \pm0.49\d.}
		
	\end{figure}\l{fig7}
	\begin{figure}[H]
		\centering
		\includegraphics[width=6cm,height=6cm,angle=0]{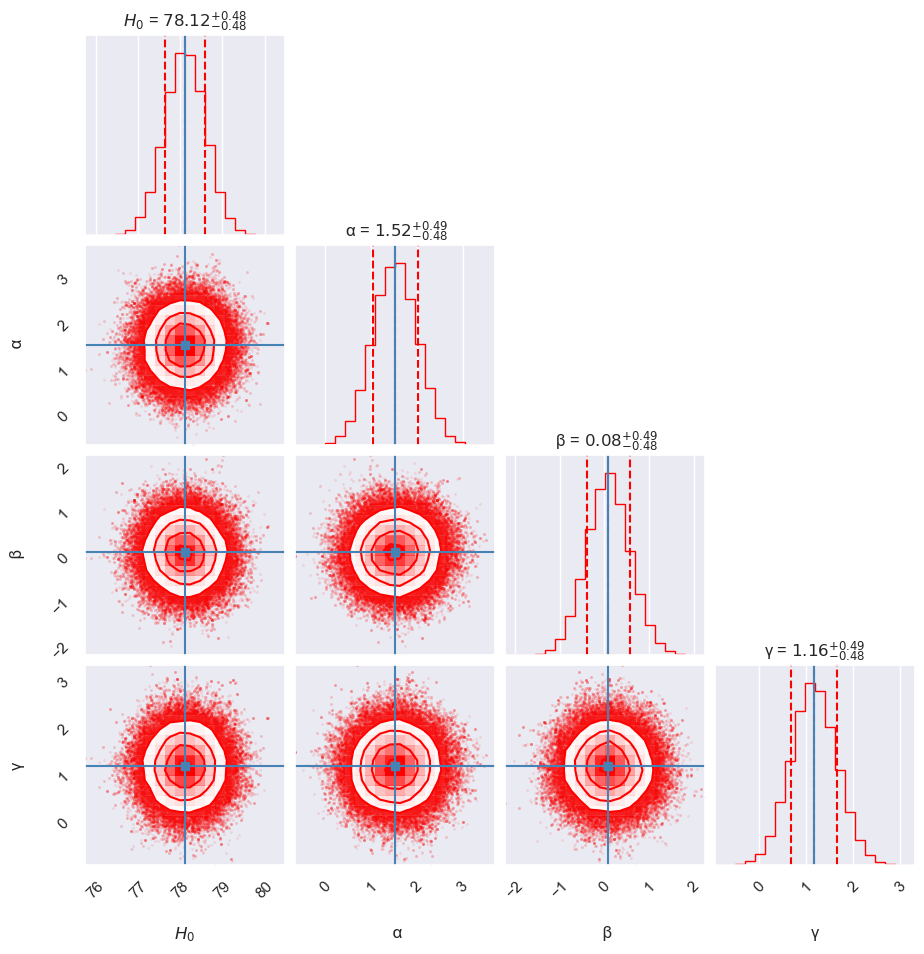}\caption{Markov Chain Monte Carlo (MCMC) based estimations for Model
			parameters $\a$, $\b$, $\c$ and $H_{0}$ and Corner plots for
			1048 Pantheon data set.}
	\end{figure}\l{fig8}

	\subsection{Baryon Acoustic Oscillations Analysis:}\l{D} BAO
	refers to the regular, periodic fluctuations in the density of
	visible baryonic matter (normal matter) of the universe, caused
	by acoustic waves in the early universe. BAO provides a
	"standard ruler" for length scale in cosmology. The scale of BAO
	is determined by the sound horizon at the time of recombination.
	By measuring the scale of BAO in the distribution of galaxies or
	in the Lyman-alpha forest of quasars, astronomers can determine
	distances across different epochs of the universe. BAO
	measurements help to trace the expansion history of the
	universe, complementing the information obtained from SNIa and
	CMB.
	\subsubsection{Key Parameters in BAO Analysis}
	\textbf{Sound Horizon ( $ r_s(z) $):} The distance that sound
	waves could travel in the early universe before recombination.
	This serves as the standard ruler for BAO measurements.It is
	defined as $$r_s(a) = \int_0^a \frac{c_s da}{ a^2 H(a) }. $$
	It's typically about 150 mega parsecs (Mpc).\\
	
	\textbf{Angular Diameter Distance ( $ d_A(z) $):} The distance
	derived from the angular size of BAO features at a given
	redshift. It is defined as
	$$  d_A(z) = c \int_0^{z} \frac{dz'}{H(z')} $$.
	\textbf{Volume-Averaged Distance ($ D_v(z) $):} A combination of
	the angular diameter distance and the Hubble parameter, given
	by: $$D_v(z) = \bigg(\frac{z {d_A}^2 (z)}{ H(z)}
	\bigg)^{\frac{1}{3}} $$
	\textbf{Ratio of Distances:} The distance redshifts ratio is
	often used in BAO analysis to compare with theoretical
	predictions. It is given by:
	\begin{equation}
		d_z(z)= \frac{r_s(z*)}{D_v(z)}
	\end{equation}
	where $ r_s(z*) $ denotes the co-moving sound horizon at the
	time when photons decouple and $ z* $ is the photons decoupling
	redshift. We consider $ z* = 1090 $ for the analysis.
	In this work, we consider a sample of BAO distance measurements
	from different surveys such as SDSS(R)\cite{Padmanabhan:2012hf}, the
	6d F Galaxy survey \cite{Beutler:2011hx}, BOSS CMASS \cite{BOSS:2013rlg},
	and three parallel measurements from the Wiggle Z survey
	\cite{Blake:2012pj,Blake:2011en,SDSS:2009ocz,SDSS:2005xqv}.\\
	
	We consider the following observational data set for our
	analysis which are described as follows \cite{ref26}:\\
	
	\d z_{BAO} \d = ([0.106, 0.2, 0.35, 0.44, 0.6, 0.73])\\
	\d d_z(z_{BAO})\d  = ([30.95, 17.55, 10.11, 8.44, 6.69, 5.45])\\	\d\sigma\d  = ([1.46, 0.60, 0.37, 0.67, 0.33, 0.31])\\
	
	The $ \chi^{2}_{BAO} $ corresponding to BAO measurements is
	given by \cite{ref26}
	\begin{equation}
		\chi_{BAO}^2(\a,\b,\c,H_0) = X^{T} C^{-1} X,
	\end{equation}
	where,
	\begin{equation}
		X = 
		\begin{bmatrix}
			$$ \frac{d_A(z*)}{D_v(0.106)}-30.84$$ \\
			$$ \frac{d_A(z*)}{D_v(0.35)}- 10.33 $$\\
			$$ \frac{d_A(z*)}{D_v(0.57)}- 6.72 $$\\
			$$ \frac{d_A(z*)}{D_v(0.44)}- 8.41 $$\\
			$$ \frac{d_A(z*)}{D_v(0.6)}- 6.66 $$\\
			$$ \frac{d_A(z*)}{D_v(0.73)}- 5.43 $$
		\end{bmatrix}
	\end{equation}
	
	and $ C^{-1} $ is the inverse of the covariance matrix given by \cite{ref26}
	
\begin{equation}\l{15}
	C^{-1} = 
	\begin{bmatrix}
		0.52552 & -0.03548 & -0.07733 & -0.00167 & -0.00532 & -0.0059 \\
		-0.03548 & 24.9707 & -1.25461 & -0.02704 & -0.08633 & -0.09579 \\
		-0.07733 & -1.25461 & 82.9295 & -0.05895 & -0.18819 & -0.20881 \\
		-0.00167 & -0.02704 & -0.05895 & 2.9115 & -2.98873 & 1.43206 \\
		-0.00532 & -0.08633 & -0.18819 & -2.98873 & 15.9683 & -7.70636 \\
		-0.0059 & -0.09579 & -0.20881 & 1.43206 & -7.70636 & 15.2814 \\
	\end{bmatrix}
\end{equation}

	We explain that the ratio \g\d d_z(z)\d\g is in fact function of
	model parameters \g\d \a,\b,\c, \g \and \g H_0\d\g. The same is
	true for $\chi^2$ function also. So we use it to estimate
	our model parameters as we did in previous sections. We presents our results in the form of following table-5 and figures 10, 11 \& 12.
	\begin{table}[H] 
		\centering
		{\begin{tabular}{c|c|c|c|}
				
				\hline\hline 
				Parameters & Least square & Least\g\d \chi^2\d\g & MCMC
				simulation\\
				& Estimations & Estimations & Estimations\\\\
				\hline\\
				\d\a\d & $ 1.456$ & $1.46\pm  0.020$ & $1.454^ {+0.974}_{-0.992} $\\\\			\hline\\
				\d\b\d & $ 1.556$ & $  1.56\pm 0.020$ & $ 1.574^{+0.960}_{-1.003}$\\\\			\hline\\
				\d\c\d & $1.899 $ & $ 1.90\pm 0.01$ & $ 1.920^{+0.582}_{-0.627}$\\\\
				\hline\\
				\d H_{0}\d & $71.386$ & $ 71.39\pm 1$ & $ 71.318 ^{+2.473} _
				{-2.283}$\\\\
				
				\hline\hline
		\end{tabular}}
		\caption{\small{ The best fit values of the model parameters
				$\a$, $\b$, $\c$ and $H_{0}$ for the best fit distance ratio
				\g\d d_{z}\d\g curve.}}
	\end{table}\l{table5}
	\begin{figure}[H]
		\centering
		
		a.\includegraphics[width=6cm,height=6cm,angle=0]{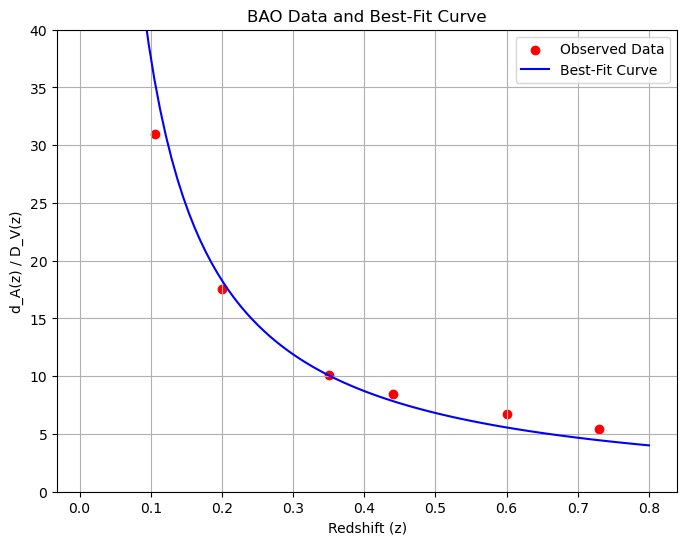}
		b.\includegraphics[width=6cm,height=6cm,angle=0]{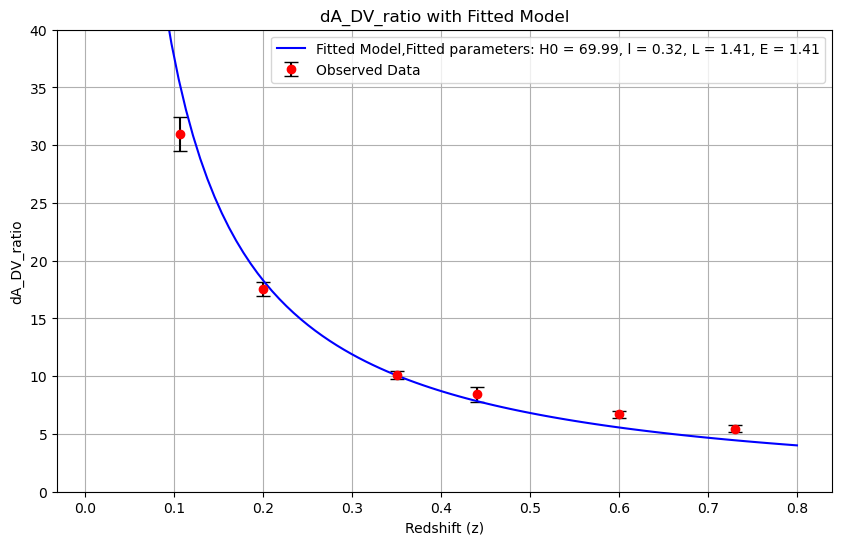}
		\caption{The distance ratio \g\d d_{z}\d\g versus redshifts best
			fit plots on the basis of least square and minimum\g \d \chi^2 \d
			\g function value . Estimated values are \d H_0 = 65 \pm 1, \a =
			0.11\pm 0.020, \b = 0.41\pm 0.020,\c = 0.91\pm 0.01\d.}
		
	\end{figure}\l{fig10}
	\begin{figure}[H]
		\centering
		\includegraphics[width=6cm,height=6cm,angle=0]{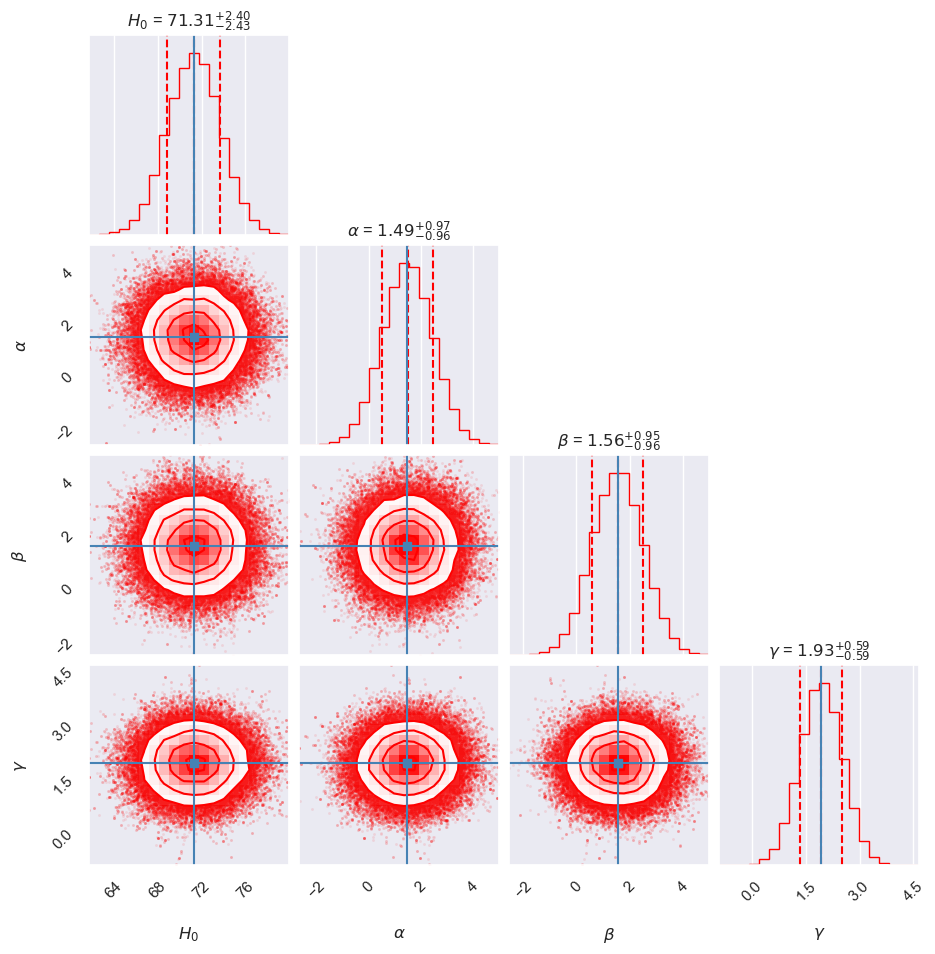}
		\caption{Markov Chain Monte Carlo (MCMC) Simulation based estimations of Model
			parameters $\a$, $\b$, $\c$ and $H_{0}$ and Corner plots for
			distance ratio \g\d d_{z}\d\g function.}
	\end{figure}\l{fig11}
	

	\subsection{Estimations of Model parameters from the combined
		data sets out of OHD, BAO, Pantheon and Union2.1 Compilation
		:}\l{E}
	In this section, we continue our estimations of model parameters
	$ a$, $\b$, $\c$ and $H_{0}$ with the help of combined data sets
	formed out of OHD, BAO, Pantheon and Union2.1 Compilation. For
	this we construct a combined $\chi^{2}$ function by adding
	individual $\chi^{2}$ function. For example suppose we want to
	combine OHD, BAO and Pantheon data sets then our combined chi
	square function will be as follows:
	\be
	\chi^2_{OHD+PAN+BAO}(\a, \b, \c, H_{0} ) = \chi^2_{OHD}(\a, \b,
	\c, H_{0} )+ \chi^2_{PAN}(\a, \b, \c, H_{0} )+ \chi^2_{BAO}(\a,
	\b, \c, H_{0} )
	\ee
	By minimizing the above function on giving proper initial values
	and ranges of model parameters, we estimate model parameters for
	combined data sets(CDS). We present the following table (VI)
	which displays estimated results for various CDS.
	
	\begin{table}[H]
		\begin{center}
			\begin{tabular}{l c c c c c} 
				\hline
				\hline 
				
				\\ 
				Parameters~~ &~~OHD+UNION ~~~~& ~~OHD+BAO+Union &
				~OHD+Pan+BAO+Union
				\\
				\hline
				\\  
				
				$H_0$ ~~&~~ $67.4^{+0.206}_{-0.205} $ & $
				67.427^{+0.197}_{-0.199} $ & $ 74.997^{+0.143}_{-0.145}$\\
				\\
				\hline
				\\ 
				$ \alpha $ & $ 0.973\pm 0.019$ & $0.973\pm 0.013 $ & $1.364\pm
				0.011 $
				\\\\
				\hline
				\\
				$ \beta $ & $ 0.9 \pm 0.019 $ & $ 0.901 \pm 0.014 $ & $ 0.885
				\pm0.012$
				\\\\
				\hline
				\\
				$\gamma$ ~~&~~$ 0.673\pm 0.006$& $0.676 \pm 0.005 $ &
				$1.01739356 \pm 0.001$
				\\
				\\
				\hline\hline   
			\end{tabular}  
			\caption{Estimated values of Model Parameters $ \alpha $, $\beta $, $\gamma$ and $ H_0 $ }
		\end{center}
	\end{table}\label{table6}
	Note that the above estimated values are based on carrying
	Morkov Chain Monte Carlo simulations which further refine the
	estimations by minimum $\chi^{2}$.
	Now, we present the following Corner plots and Step number plots (Fig. 13)
	to show the results of MCMC simulations. The details have been
	given in the caption of each plot.
	
	
	\begin{figure}[H]
		\centering
		\includegraphics[width=4cm,height=6cm,angle=0]{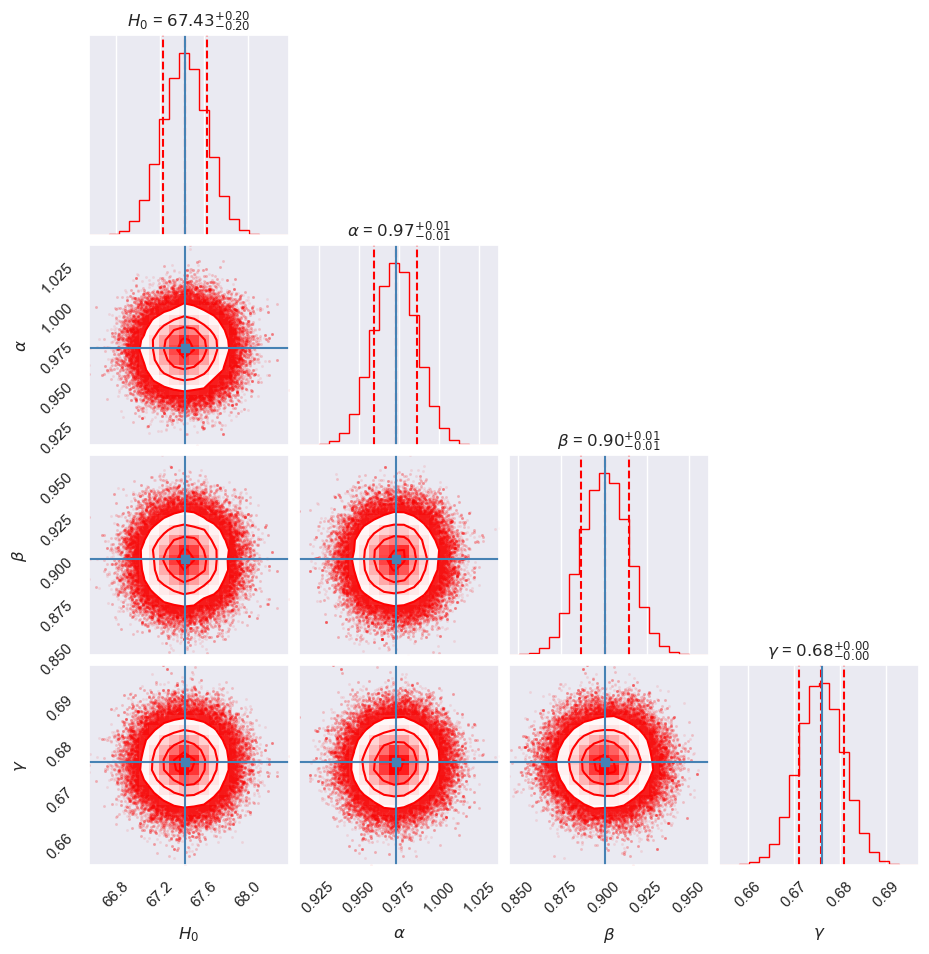}
		\includegraphics[width=4cm,height=6cm,angle=0]{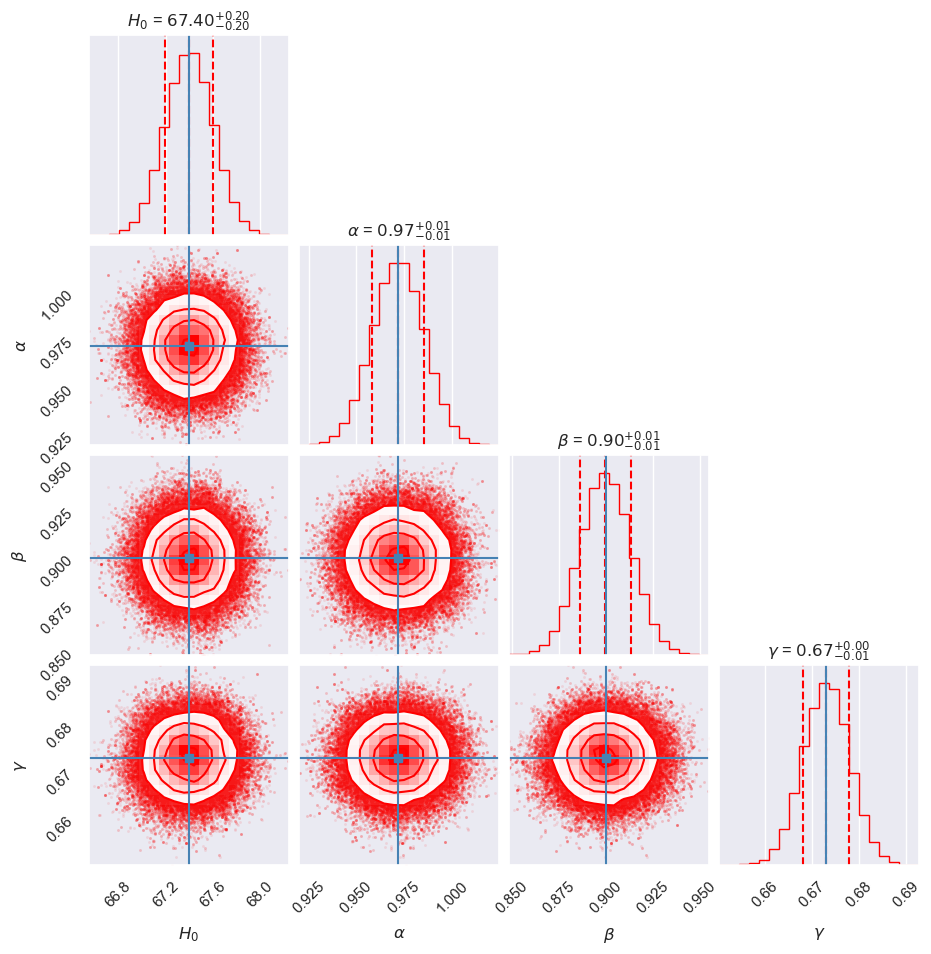}
		\includegraphics[width=4cm,height=6cm,angle=0]{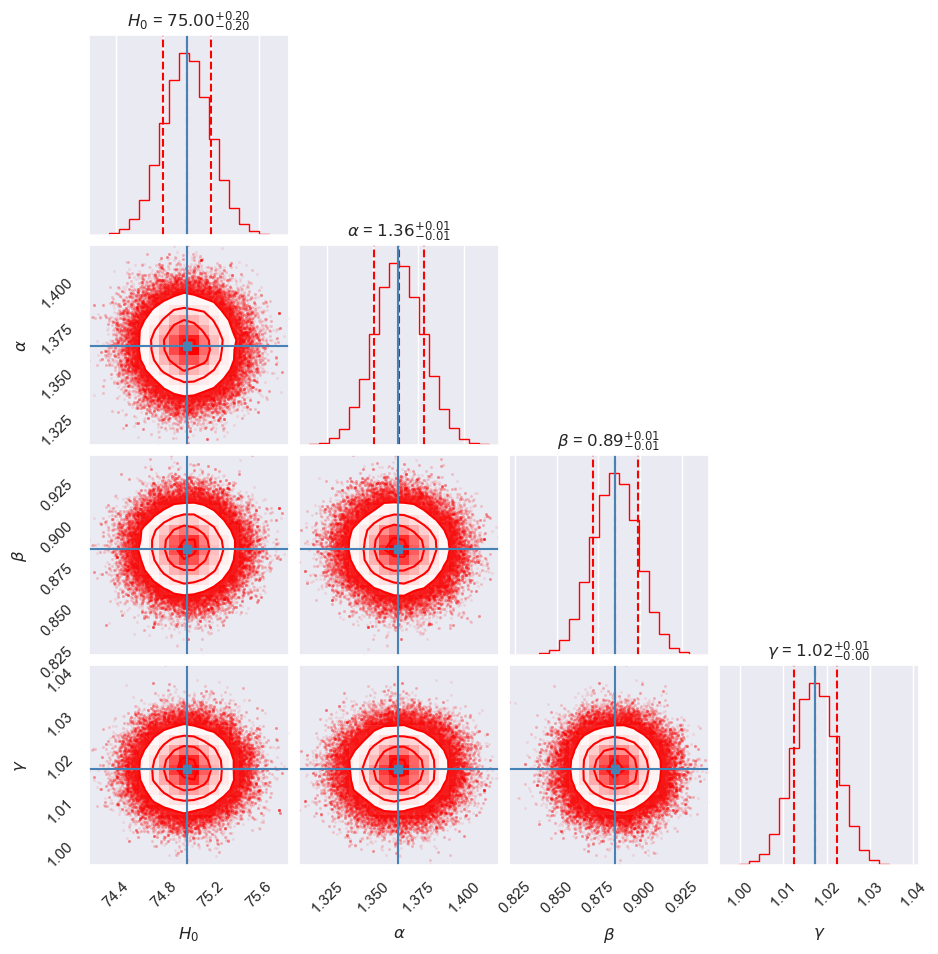}
		\includegraphics[width=4cm,height=6cm,angle=0]{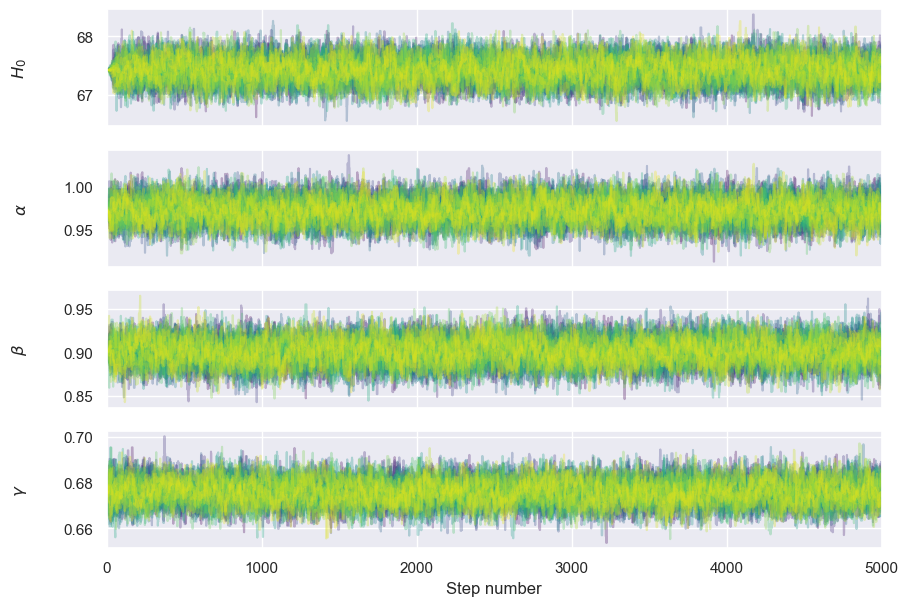} 
		\caption{\small{Corner plots for Combined Hubble plus BAO plus
				Union 2.1 compilation and Hubble plus Union2.1 compilation data
				sets, based on Markov Chain Monte Carlo (MCMC) Simulation
				estimations of Model parameters $\a$, $\b$, $\c$ and $H_{0}$.}
			The last two figures are Step numbers versus Model parameters
			$\a$, $\b$, $\c$ and $H_{0}$ plots for combined data sets
			described earlier. }
	\end{figure}\l{fig13}
	\section{Hubble Tension}
Various techniques have been employed in the 21st century to ascertain the Hubble constant. Utilizing calibrated distance ladder techniques, ``late universe" observations have come to a consensus on a value of roughly 73 (km/s)/Mpc. A value of around 67.7 (km/s)/Mpc is agreed upon by ``early universe" methodologies that rely on cosmic microwave background observations, which have been accessible since 2000. The investigation \cite{102} is similar to the first number since it takes into account the variation in the expansion rate since the early universe. The estimated measurement errors have decreased with improved methodologies, but the measured value range has not decreased, to the point that the disagreement is now extremely statistically significant. The ``Hubble tension" refers to this disparity \cite{103,104}.\\
	We furnish the following list of present values of Hubble
	parameter\g$H_0$\g obtained from latest surveys and projects.

	\begin{table}[H]
		\begin{center}
			\begin{tabular}{ |c| c |c |c |c |c } 
				\hline
				\hline 
				\\ 
				Date published~&~Hubble constant & ~Observer ~ &~ Citation
				&~methodology \\
				\\
				& (km/s)/Mpc    &               &               &\\
				\hline
				\\  
				2022-02-08 & $73.4^{+0.99}_{-1.22}$ & Pantheon+ & \cite{97} & ``SN Ia distance ladder (+SHOES)"\\
				\\
				\hline
				\\ 			
				2022-06-17 & $75.4^{+3.8}_{-3.7}$ & T. de Jaeger et al. & \cite{98} &
				\small{ ``Type II supernovae as standardisable candles"}\\
				\\
				\hline
				\\			
				2021-12-08 & $ 73.04 \pm 1.04 $ & SHOES & \cite{99} &
				\small{``Cepheids-SN Ia distance ladder" (HST+Gaia
					EDR3+"Pantheon+")}.\\
				\\
				\hline
				\\	
				2023-07-19 & $67.0\pm3.6$ & Sneppen et al. & \cite{96} &
				``Blackbody spectra of the optical counterpart"\\
				& & & & of neutron-star mergers,\\
				\\
				\hline
				\\
                2022-12-14& $67.3^{+10.0}_{-9.1}$ & S. Contarini et al. & \cite{100} &
				``Statistics of cosmic voids using BOSS DR12 data set"\\
				\\
				\hline
				\\
				2023-07-13 & $68.3\pm1.5$ & SPT-3G & \cite{101} & ``CMB TT/TE/EE power
				spectrum".\\
				\\
				\hline\hline  
			\end{tabular}  
			\caption{ Latest $ H_0 $ Related Findings on the basis of
				Various Empirical Experiments and Observations}
		\end{center}
	\end{table}\label{table7}
	If we compare results of tables \g$6$\g with those of
	table\g$7$\g, we find that our estimations of Hubble parameter
	\g$H_0$\g are very much close to the observed values.
	\subsection{Present Density and the Present Cosmological
		constant:}
	We recall the following Equations for density of the universe
	and cosmological constant:
	
	\be \no
	\rho(z) = \rho_c \frac{2(\alpha +\beta -3) (z+1)^{\frac{2 (w+1)
				(\alpha +\beta -3)}{-\gamma +\alpha (w+1)-2}}}{3 (\alpha -\gamma
		+\alpha w-2)}
	\ee
	and
	\be\no
	\Lambda(z)=\frac{{H_0}^2 (\alpha -2 \beta -3 \gamma +3 \alpha w)
		(z+1)^{\frac{2 (w+1) (\alpha +\beta -3)}{-\gamma +\alpha
				(w+1)-2}}}{\alpha -\gamma +\alpha w-2},
	\ee
	where $\rho_c= \frac{3H_0^2}{8\pi G}$ is the critical density of
	the universe.

	From these equations and using tables (VI), we obtain the
	present values of density of the universe and that of
	cosmological constant. We also find Vacuum Energy of the
	universe which is defined as:
	\be\no
	\rho_{\Lambda}= \frac{\Lambda c^2}{8 \pi G}.
	\ee
	We note that $ \frac{3c^2}{8 \pi G} =0.000189*10^{29} g/cm^3 .$
	\\
	The results are described in the following table (8).
	
	\begin{table}[H]
		\begin{center}
			\begin{tabular}{l c c c c c} 
				\hline
				\hline 
				\\ 
				
				Parameters~~ ~& ~~OHD+BAO+Union  &  ~OHD+Pan+BAO+Union  ~\\
				\\
				\hline
				\\  
				
				$H_0$ ~ & $ 67.427 $ &  $ 74.997$\\
				\\
				\hline
				\\ 
				$ \Omega_m $  & $0.435453 $ & $0.305221$
				\\\\
				\hline
				\\ 
				$ \Omega_{\Lambda} $ & $ 0.564547 $& $0.694779$
				\\\\
				\hline
				\\
				$\rho_0 $ & $ 0.373872*10^{-29}gm/cm^3 $&
				$0.324389*10^{-29}gm/cm^3$
				\\\\
				\hline
				\\
				$\Lambda_0$ ~~ & $8.06442*10^{-32} sec^{-2}$ & $12.2783*10^{-32}
				sec^{-2}$
				\\\\
				\hline
				\\
				$ \rho_{\Lambda_0} $ ~~&~~$4.85098*10^{-30}gm/cm^3$ &
				$7.38578*10^{-30}gm/cm^3$
				\\\\
				
				\hline\hline 
			\end{tabular}  
			\caption{Estimated values of Density of the Universe and
				Cosmological Constant }
		\end{center}
	\end{table}\label{table8}
	We also present the plot (Fig. 14) to describe variation of density
	over redshifts which is a increasing one, This shows that in the
	past the density of the universe was more and it is decreasing
	over time. We note that the redshift is proportion to the time.
	\begin{figure}[H]
		\centering
		\includegraphics[width=10cm,height=6cm,angle=0]{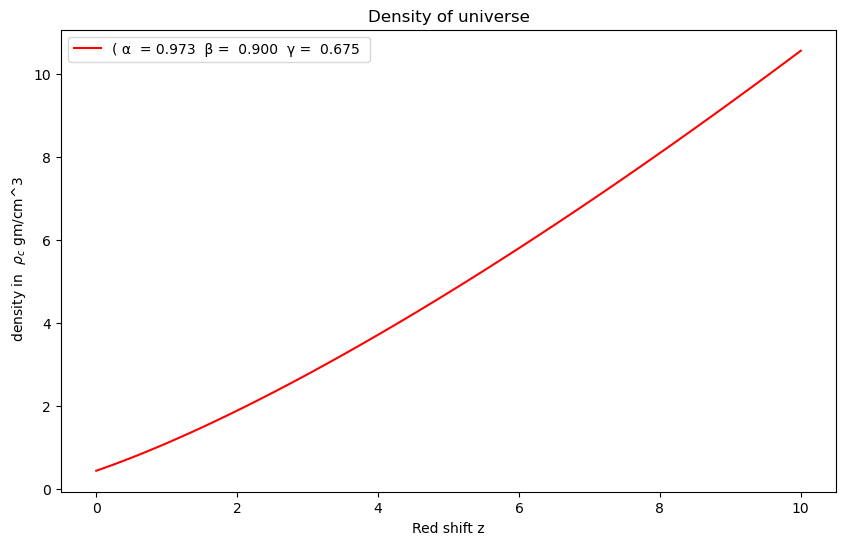}
		\caption{ Density versus redshift Plot }
	\end{figure}\l{fig14}
	\subsection{Conversion of Redshift into Time and Age of the
		Universe}
	We may get time in billion yrs in term of redshift from
	the following relation :
	\be\no
	\frac{dz}{dt} = -(1+z)H(z)
	\ee
	This equation is integrated to yield the following relation ship
	between elapsed time\g$(t_0- t_z$)\g from present and redshift
	$z$:
	\be
	(t_0 - t_z) = \int_0^z\frac{ dz}{(1+z)H(z)}=\frac{\small
		(\c-\a+2) }{ (-3+\a+\b)H_0 }((1+z)^{\frac{ -3+\a+\b}{\c -\a
			+2}}-1)
	\ee\l{time}
	where $t_0$ is present time at $z=0$ and $t_z$ is the time at
	redshift $z$, so that $(t_0 -t_z$) will be the elapsed time
	from present at the redshift $z$.
	We note that unit of Hubble parameter $H(z)$ \small{ Km/sec/Mps}
	which is in fact a unit of reciprocal of time. We use following
	conversion formula to express $H(z)$ into reciprocal of billion
	yrs.
	\be\no
	\frac{Mps}{km/sec} = \small{976.32}\g \small{billion yrs}.
	\ee
	Now we present the following plot (Fig. 15) to show graphically redshift
	versus time relation:
	\begin{figure}[H]
		\centering
		\includegraphics[width=10cm,height=6cm,angle=0]{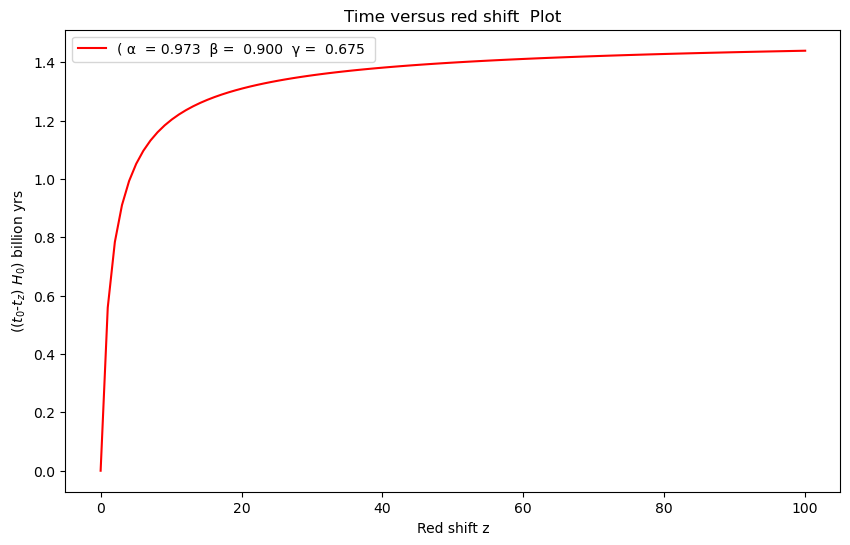}
		\caption{ Time versus redshift Plot }
	\end{figure}\l{fig15}
	From this, we observe that there is an asymptote at $z=1.4998.$
	This gives us age of the universe which is described in the
	table (9) as follows:
	\begin{table}[H]
		\begin{center}
			\begin{tabular}{l c c c c c} 
				\hline
				\hline 
				
				\\ 
				Parameters~~ ~& ~~OHD+BAO+Union & ~OHD+Pan+BAO+Union 
				\\
				\hline
				\\  
				
				$H_0$ ~ & $ 67.427^{+0.197}_{-0.199} $ & $
				74.997^{+0.143}_{-0.145}$\\
				\\
				\hline
				\\ 
				$ \alpha $   & $0.973\pm 0.013 $ &  $1.364\pm 0.011  $
				\\\\
				\hline
				\\
				$ \beta $  & $ 0.901  \pm 0.014 $ &  $ 0.885 \pm0.012$ 
				\\\\
				\hline
				\\
				$\gamma$ & $0.676 \pm 0.005 $ &  $1.017 \pm 0.001$
				\\\\
				\hline
				\\
				
				Age of The Universe & 21.7282 billion yrs & 22.7808 billion yrs
				\\\\
				\hline\hline   
			\end{tabular}  
			\caption { Age of the Universe as per estimations of model
				parameters }
		\end{center}
	\end{table}\label{table9}
\section{Conclusions} We developed a universe models of the FLRW space time that is associated with a variable cosmological term $\Lambda(t)$. For this the functional form of $\Lambda(t)$ is taken as $\a\frac{\ddot{a}}{a} +\b(\frac{\dot{a}}{a})^2 +\c \rho$. We have been able to derive a power law solution for Hubble parameter H(z), cosmological constant $\Lambda(z)$ and densities(matter and Lambda vacuum) of the universe in term of redshifts z and our model turns out to be a Einstein-de Sitter accelerating one. There are four model parameters $ H_0, \a, \b, \and \c $ which are were estimated on the basis of the four data sets. The Hubble 46 data set describing Hubble parameter values at various redshifts, Union 2.1 compilation data sets comprising of distance modulus of 580 SNIa supernovae at different redshifts, The Pantheon  data set which contains Apparent magnitudes of 1048 SNIa supernovae at various redshifts and finally BAO data set of  volume averaged distances at 5 redshifts. We employ the conventional Bayesian methodology to analyze the observational data and also the Markov Chain Monte Carlo (MCMC) technique to derive the posterior distributions of the parameters. For MCMC analysis, we utilize the $ emcee $ package to determine the best-fit values of the model parameters. Along with this, we  have also used technique of minimizing $\chi^{2}$ function for parameter estimation. As a conclusion, we present our findings item wise as follows:
\begin{itemize}
\item we have carried out a detailed analysis of the universe expansion. Our investigation applies that the Hubble function envisages an increasing trend over redshift, which means that in the past the Hubble parameter was carrying high value and over the time it is gradually decreasing (See figs 1(a)and  1(b)). The  best fit values of Hubble parameter $H_0$ as per the four data sets described earlier are found as $61.53^{+0.453}_{-0.456}$, $ 69.270^{+0.229}_{-0.228}$, $78.116^ {+0.480}_{-0.479}$ and $ 71.318 ^{+2.473}_{-2.283}$ respectively.
\item Off late  the present value of Hubble parameters $H_0$ were empirically given as 73 and 67.7 (km/s)/Mpc using  distance ladder techniques and measurements of the cosmic microwave background. We also tried to achieve these values statistically by using combined data sets out of the four described earlier. The OHD+BAO+Union and ~OHD+Pan+BAO+Union combined data sets provides the best fit Hubble parameter value $H_0$ as 	$67.427^{+0.197}_{-0.199}$ and $74.997^{+0.143}_{-0.145}$ respectively.
\item We also worked out in finding the  growth of matter and vacuum cosmological constant related energy densities of the universe. They do follow the increasing trend  over redshift which interprets that in the past the densities were high and over the time they are gradually  decreasing due to the expansion of the universe. We have estimated that the present values  $\rho_0 $ and $\rho_{\Lambda_0} $ of matter and Lambda vacuum densities are $3.244*10^{-30}gm/cm^3$ and $7.39*10^{-30}gm/cm^3$ respectively. The higher value of vacuum energy shows the dominance of this energy at present. We recall that this energy is responsible for the present acceleration of the universe as it creates negative anti gravitating pressure in the universe.
\item Lastly, we have calculated the universe's current age and established the relationship between time and redshift. Our model estimates the current age of the universe to be 21.7282 billion years, which is significantly more than the $\Lambda$-CDM model's prediction.
\item In summary, the dynamical cosmological constant is a crucial concept in exploring the deeper nature of dark energy and how it governs the expansion and ultimate fate of the universe. It offers a more flexible framework than a static $\Lambda$, potentially solving several outstanding cosmological problems.

\end{itemize}
\section*{Declaration of competing interest}
The authors declare that they have no known competing financial interests or personal relationships that could have appeared to influence the work reported in this paper.

\section*{Data availability}
No data was used for the research described in the article.

\section*{Acknowledgments}
The IUCAA, Pune, India, is acknowledged by the author ( A. Pradhan) for giving the facility through the Visiting Associateship programmes. 

\section*{ORCID}
Anirudh Pradhan: https://orcid.org/0000-0002-1932-8431 \\
G. K. Goswami: https://orcid.org/0000-0002-2178-6925

\end{document}